\documentclass[12pt]{article}
\usepackage{geometry}
  \usepackage{graphicx}
  \usepackage{amsmath, amsthm, amsfonts}
  \usepackage{color,soul}
  \usepackage{hyperref}

\begin{document}


\title{Focused Ion Beam fabrication of Janus bimetallic cylinders acting as drift~tube Zernike~phase~plates} 
\maketitle

\author{Paolo~Rosi$^1$,Gian~Carlo~Gazzadi$^2$,Stefano~Frabboni$^{1,2}$, Vincenzo~Grillo$^2$,Amir~H.~Tavabi$^3$, Rafal~E.~Dunin-Borkowski$^3$, Giulio~Pozzi$^{3,4}$}
\null
\noindent
$^1$ \small Department FIM, University of Modena and Reggio Emilia,  via G. Campi 213/a, Modena, 41125, Italy \\
$^2$ \small Department of Physics, University of Ottawa, 25 Templeton Street, Ottawa, Ontario K1N 6N5, Canada\\
$^3$ \small Ernst Ruska-Centre for Microscopy and Spectroscopy with Electrons and  Peter Gr\"unberg Institute, Forschungzentrum~J\"ulich,~52425~J\"ulich,~Germany\\
$^4$ \small Department of  Physics and Astronomy, University of Bologna, viale B. Pichat 6/2, 40127 Bologna, Italy\\

\begin{abstract}
Modern nanotechnology techniques offer new opportunities for fabricating structures and devices at the micron and sub-micron level.
Here, we use  focused ion beam techniques to realize drift tube Zernike phase plates for electrons, whose operation is based on the presence of contact potentials in Janus bimetallic cylinders, in a similar manner to the electrostatic Aharonov-Bohm effect in bimetallic wires.
We use electron Fraunhofer interference to demonstrate that such bimetallic pillar structures introduce phase shifts that can be tuned to desired values by varying their dimensions, in particular their heights.
\end{abstract}



Developments in state-of-the-art electron microscopes (including aberration correctors, field emission guns and single particle detectors), in the fabrication of sub-micron devices (using focused ion beam instruments, electron beam and optical lithography, \emph{etc.}) and in biological specimen preparation (\emph{e.g.}, the use of frozen-hydrated specimens in cryo-electron microscopy) have stimulated renewed interest in using phase plates as devices that can be used to apply phase shifts to scattered electrons with respect to unscattered beams, in order to improve contrast in images of weak phase objects (including unstained biological molecules) more efficiently than using standard methods, such as defocusing the objective lens.

A survey of  the history of phase plates has been presented by Nagayama, \cite{Nagayama:2011} while the state of the art in the field has been reported by Glaeser,\cite{Glaeser:2013} who summarised the strengths of each device, highlighted remaining problems and presented future perspectives.
In his analysis, he listed electrostatic drift tubes, \emph{i.e.}, electrically biased drift tubes surrounded by grounded guard electrodes,\cite{Cambie:2007} as a promising approach.
One of the problems with fabricating such devices is the need to connect and control the electrical bias to each nanoscale electrode, with limited space available for cabling.
Here, we show how similar devices  can be realized by making use of contact potential differences between metals to generate potential differences. This concept loses the advantage of tuneability, but is simpler in realization and offers prospects for further reductions in device dimensions.
A desired constant phase shift can be chosen by selecting the dimensions of such a device, in particular its height.
A similar use of contact potentials has been reported by two groups\cite{Perry:2012,Tamaki:2013} for a two-metal Einzel lens configuration (Boersch phase plate),\cite{Glaeser:2013} which has a different configuration from the present ``drift tube'' concept.

\begin{figure}[ht]  \centering \includegraphics[width=7cm]{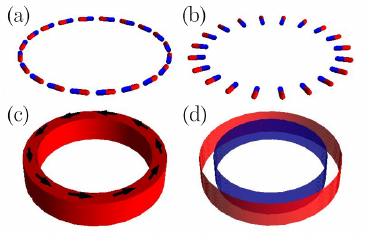} 
\caption{Analogies between arrangements of magnetic (left) and electric (right) dipoles. (a)~Toroidal magnetic flux tube and (b)~toroidal ring of electrostatic dipoles, represented by discrete sets  of dipoles. (c)~Toroidal magnet and (d)~drift tube formed from coaxial cylinders of opposite charge.}   
\label{scheme}  \end{figure}

The working principle of the proposed device can be understood within the framework of the  Aharonov-Bohm effect \cite{Aharonov:1959,Aharonov:1984,Olariu:1985} in its electrostatic version,\cite{Boyer:1973,Boyer:2002} as demonstrated previously for a bimetallic wire.\cite{Matteucci:1982a,Matteucci:1985,Matteucci:1992b,Matteucci:2002}
It is straightforward to demonstrate the  equivalence of the electron optical phase shifts produced by magnetic and electrostatic dipoles that are rotated with respect to each other by $90^\circ$. \cite{Pozzi:2016}
For a toroidal instead of a linear geometry, a circular distribution of elementary magnetic dipoles, which is shown in discrete form in Fig.~\ref{scheme}(a), is equivalent to a circular distribution of electrostatic dipoles, as shown in Fig.~\ref{scheme}(b).
Each arrangement produces a constant phase shift between an electron beam passing inside and outside the toroid.
On a macroscopic level, this equivalence is retained between a toroidal magnet, as shown in Fig.~\ref{scheme}(c), and a cylindrical distribution of charge that has a radial electric dipole moment, such as the drift tube shown in Fig.~\ref{scheme}(d).

In order to test this proposal, we fabricated Janus bimetallic cylinders and measured their phase shifts with respect to nearby holes of nearly equal diameter.
We recorded Fraunhofer images of each pair of holes, from which the phase shift of the drift tube with respect to the nearby hole could be measured in the form of a lateral displacement of the interference fringes with respect to the diffraction envelope.\cite{Venturi:2017} This effect was enhanced by optimizing the  radii of the holes and their separations. \cite{Born:1969,Pozzi:2016}

 A rough estimate of the phase difference can be obtained by assuming that the potential inside a cylinder of height $h$ is constant and equal to the contact potential difference $V_C$ between the two dissimilar metals, in our case Pt and Al, which is $\sim 1V $.\cite{Michaelson:1977,Rumble:2020}
The corresponding phase difference is  given by the expression \cite{Kohl:2008,Pozzi:2014,Pozzi:2016}  
   \begin{equation}
 \phi = C_E V_C  h~,
 \label{phase}
 \end{equation}
 where $C_E$ is an interaction constant that takes a value of $7.3 {\rm  rad}/({\rm V \mu m})$ for $ \rm 200~keV$ electrons.
 The expected phase shift is $2 \pi$ for a tube length $h \sim 0.86~\mu$m.

\begin{figure}[ht]  \centering \includegraphics[width=7cm]{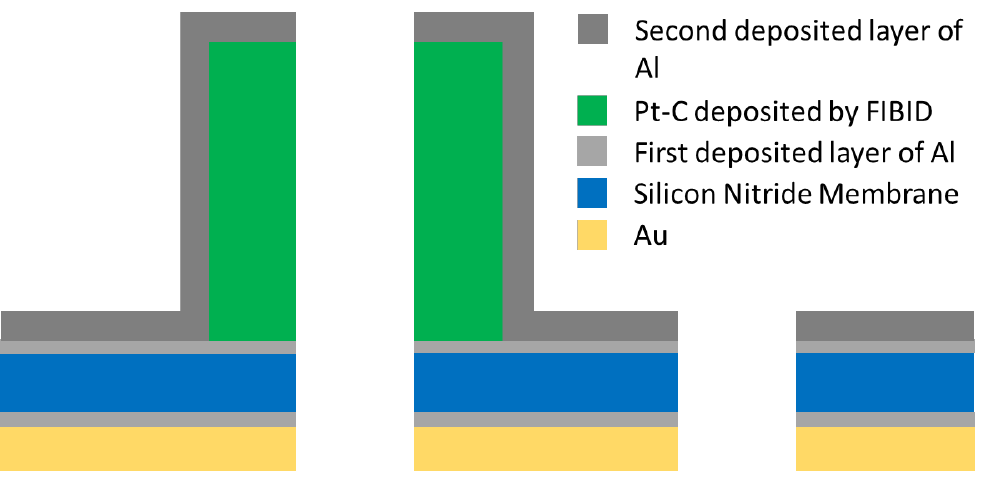} 
\caption{Schematic diagram of the cross-sectional structures of the cylinders and adjacent holes.}   
\label{scheme2}  \end{figure}

\begin{figure}[ht]  \centering \includegraphics[width=8cm]{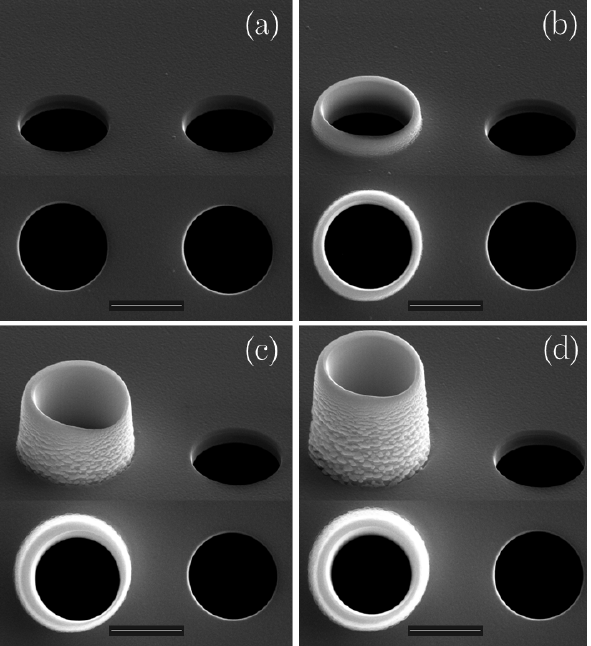} 
\caption{ SEM tilted and top views of the double circular aperture structures. One aperture in each pair takes the form of a FIBID-fabricated Al-covered Pt-C tube. Several representative values of tube height are shown. The scale bar in each image is $2~\rm\mu m$. The measured pillar heights are (a)~0 (\emph{i.e.}, no pillar structure), (b)~0.7,  (c)~2.2 and (d)~3.3~$\mu$m.} 
\label{pillars}  \end{figure}

We used focused ion beam (FIB) techniques to realize a toroidal version of a bimetallic wire.
The starting point was a commercial silicon nitride membrane of thickness $200~\rm nm$, with nine $100\times100~\rm\mu m$ windows, onto both sides of which $\sim 50~\rm nm$ of Al was evaporated.
Pairs of circular holes, with a pillar on one hole in each pair, were fabricated in four steps to realize the cross-sectional structures shown in Fig.~\ref{scheme2}.
First, Pt-C cylinders were deposited by FIB-induced deposition (FIBID) using a Pt-organic precursor gas.
For each cylinder, the ion beam was scanned during gas injection over a ring-shaped area with an inner radius of $1.15~\rm\mu m$ and an outer radius of  $1.5~\rm \mu m$.
Five pillars of varying height were deposited.
The total scan time was $\sim1$~minute for the shortest pillar and $\sim9$~minutes for the tallest pillar (Fig.~\ref{pillars}.d). A distance of at least 35~$\rm \mu m$ was kept between pillars and no more than three pillars were deposited on each window.
In the second step, a bimetallic structure was created by depositing $\sim100~\rm nm$ of Al over the entire sample using thermal evaporation.
In the third step, a further $\sim 150~\rm nm$ of Au was deposited onto the ``lower'' side of the membrane using sputter coating, in order to make the support opaque to electrons.
In the final step, a double aperture was formed by boring pairs of holes with radii of $1.2~\rm\mu m$ using FIB milling, one through each tube and one adjacent to it.
The distances between the centers of adjacent holes was approximately twice the diameter of each hole.

Figure \ref{pillars} shows scanning electron microscopy (SEM) images recorded in tilted and top view configurations. A reference structure of two holes without a fabricated pillar structure is shown in Fig.~\ref{pillars}(a). The tilted views shown in Figs~\ref{pillars}(b-d) indicate that the pillar heights are 0.7, 2.2 and 3.3~$\mu$m, respectively.

\begin{figure}[ht]\centering \includegraphics[width = 8.5cm]{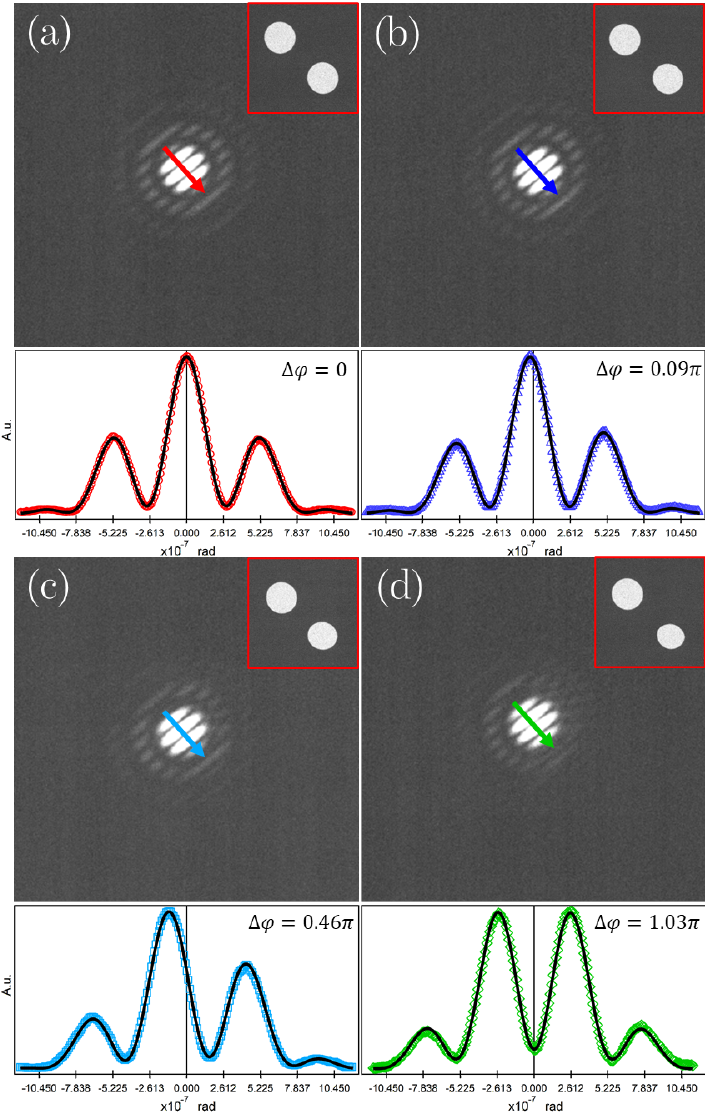}\caption{TEM images and diffraction patterns of the double apertures shown in Fig.~\ref{pillars}. Line scans of the diffraction patterns are shown, with the markers representing experimental data points and solid black lines fits. The direction of the line scan is marked by an arrow in each diffraction pattern. The phase differences (modulo~$2\pi$) obtained from fits to Eq.~\ref{fitting} are indicated in the top right corner of each line scan.}\label{TEM_results}\end{figure}

Experiments were carried out at 200~keV on a Talos X TEM in low magnification, large angle diffraction mode (camera length 1.4~km).
Results obtained from the structures shown in Fig.~\ref{pillars} are reported in Fig.~\ref{TEM_results}.  

Fraunhofer images corresponding to the zero height pillar (Fig.~\ref{TEM_results}~(a)) and the 0.7~$\mu$m pillar (Fig.~\ref{TEM_results}~(b)) are found to be symmetrical, with a central maximum with respect to the diffraction envelope, indicating that their phase difference is a multiple of  2$\pi$.
In contrast, the results shown in Fig.~\ref{TEM_results}~(c) exhibit a marked asymmetry, suggesting a detectable phase difference, while the results shown in Fig.~\ref{TEM_results} (d) have a central minimum instead of a central maximum, indicating that the phase difference is approximately $\pi$.
Values of the phase difference were measured by performing fits to the data based on the formula 
\begin{equation}
I(x)=b \frac{J_1^2(xD)}{(xD)^2}\left[1+\mu cos\left( x d+\Delta\varphi \right)\right]~,
\label{fitting}
\end{equation}
which was derived by considering the Fraunhofer images of two circular apertures. The multiplicative factor is the Fourier transform of a circular aperture (\emph{i.e.}, an Airy disk), while the interference term in square brackets includes a damping factor $\mu$, which accounts for partial coherence effects resulting from the finite dimensions of the electron source (\emph{i.e.}, spatial coherence).\cite{Born:1969}
 $J_1$ is a Bessel function of the first kind, $b$ is a fitting parameter that allows a non-normalized intensity profile to be fitted, $\Delta\varphi$ is the phase difference between the apertures and $D$ and $d$ are coefficients that are proportional to the aperture diameter and separation, respectively.
 As a result of the high coherence of the field emission gun, the influence of spatial coherence was found to be less than $10\%$ of the  overall intensity.

The fitted values of the phase difference, which are indicated in Fig.~\ref{TEM_results}, are consistent with the generation of constant phase shifts by Janus bimetallic cylindrical structures that behave as drift tubes.
More systematic experiments will be required in the future to assess the influence on the phase shift of experimental parameters associated with the fabrication process, as well as possible effects of electron beam induced charging of both the structures and their supports.
When better control of the phase shift is achieved, it may be possible to use such structures as Zernike phase plates in transmission electron microscopy, as they are more compact than tuneable drift tube devices and therefore promise to allow a greater number of spatial frequencies to contribute to phase contrast images.

Beyond their use as Zernike phase plates, we foresee the application of such bimetallic devices for different forms of electron beam shaping. 
For example, they could be used to replace electrostatic elements in applications such as spiral phase plates,\cite{Furhapter:2007,Juchtmans:2016,Juchtmans:2016a,Pozzi:2017,Tavabi:2020} Hilbert phase plates and electrostatic orbital angular momentum sorters.\cite{Pozzi:2020,Tavabi:2019}
They could also be used to replace electrostatic phase plates for conformal mapping operations.\cite{Ruffato:2020} 
A two-dimensional array of such cylinders could be used to produce an arbitrary phase landscape without the problems that come with using material-based holograms. A similar concept was recently proposed for pixelated programmable phase plates.\cite{Verbeeck:2018,Thakkar:2020}  When using bimetallic cylinders, the loss of \emph{in~situ}  programmability would be compensated by the ability to fabricate a greater number of pixels per unit of area.

\section*{Acknowledgments}
PR and SF aknowledge the staff at the CIGS facility in Modena for helping in setting up the experiments.
The research leading to these results has received funding from the European Union’s Horizon 2020 Research and Innovation Programme under Grant Agreements: 1.  Grant No 766970 Q-SORT (H2020-FETOPEN-1-2016-2017); 2. Grant No. 823717 (project ESTEEM3).
\section*{Data Availability Statement}
The data that support the findings of this study are available from the corresponding author upon reasonable request.



\end{document}